# Efficient design of a one-hour unit test for introductory physics

Benjamin O. Tayo, Pittsburg State University, Pittsburg, KS

Designing a one-hour unit test for an introductory physics class can be quite challenging. We present an efficient format of a one-hour unit test that utilizes question groups for quantitative free-response problems and a binary system (True or False) for conceptual questions. This format promotes coherence and allows the instructor to test a wide variety of topics in a one-hour exam. For the 4 semesters in which the design has been implemented, average unit test scores have increased from 76% using traditional test format, to 79% using the efficient design.

## I. Introduction

In introductory physical science courses like introductory physics, in-class tests play an important role in assessment. In introductory physical science courses where class meeting times are only one-hour long, designing a one-hour unit test that is capable of assessing the wide variety of topics in each unit, and that can be completed within an hour is quite formidable. Several instructors have tried different ways of designing a unit test such as free-response quantitative problems where students are expected to show all necessary steps involved in their calculations; multiple-choice questions; or a unit test that combines free-response quantitative problems and multiple-choice questions. Multiple choice exams have the advantage that it takes less time to grade, but do not really assess quantitative skills as students are not required to solve the problem elaborately, while presenting all the steps and stages involved in the calculation. Quantitative problems are typically long questions with several sub-parts. In this type of problems, the student is expected to show all the necessary steps involved in their calculations and are awarded partial credits for displaying different levels of problem-solving capabilities. Several studies have been performed by different investigators to study the relative advantages of multiple-choice and free-response quantitative testing designs [1-8]. Lin and Singh have shown recently that carefully designed multiple-choice questions can reflect the relative performance on the free-response quantitative questions while maintaining the benefits of ease of grading and quantitative analysis, especially if the different choices in the multiple-choice questions are weighted to reflect the different levels of understanding that students display [8].

While traditional methods such as multiple-choice and free-response quantitative test formats have their advantages and disadvantages, these methods are not quite suitable for designing a one-hour unit test. In this article, we present an efficient format of a one-hour unit test that utilizes question groups for quantitative free-response problems and a binary system (True or False) for conceptual questions. We discuss the design aspects of the one-hour unit test as well as its advantages over traditional testing formats. We also compare average test scores from traditional and the efficient design formats based on 8 semester of data.

This article is organized as follows: In section II, we discuss the efficient design of a one-hour unit test. In section III, we discuss the advantages of the design. Finally, in section IV, a short summary concludes the article.

## II. Efficient design of a one-hour unit test

In introductory physical science courses like introductory physics, in-class tests play an important role in assessment. We focus on calculus-based engineering physics (PHYS 104) which is offered every semester at Pittsburg State University. The course covers the basics of mechanics, waves, fluids, and thermodynamics. This course covers the first 20 chapters of *Physics for Scientists and Engineers*, 4th ed., by Giancoli [9]. These 20 chapters are divided into 4 units, typically 5 chapters per unit (see Appendix A for sample syllabus). A one-hour unit test is organized at the completion of units 1, 2 and 3. The final comprehensive exam is a 3 hours exam and takes place after the completion of unit 4, and typically weighted in such a way that 40 percent is from unit 4 and 60 percent from units 1, 2, and 3. We only focus on the one-hour unit test.

We now present an efficient format of a one-hour unit test that utilizes question groups for quantitative free-response problems and a binary system (True or False) for conceptual questions. In this design, a unit test is composed of 4 questions. The first 3 questions are free-response quantitative questions with 3 to 5 subparts, each question based on a central topic. The last question is a conceptual question with 10 subparts. Each subpart in the quantitative section is a True or False type question. Using a binary system instead of a multiple-choice format allows students to complete this section within a reasonable amount of time. The design allows assessment of problem-solving skills via free-response quantitative problems and conceptual skills via a binary system of True or False questions. With this design, a wide variety of topics can be included in the test, as discussed below.

In Appendices B, C, and D, we show sample one-hour unit tests to illustrate the efficiency and coherence of the design. The three unit tests are based on unit 1, unit 2, and unit 3. Material covered in unit 4 is typically tested in the comprehensive final exam.

In appendix B, we see that unit test 1 follows the same pattern in the syllabus. Question 1 has 5 subparts, and tests knowledge on measurements, uncertainty, units, and significant figures. Question 2 covers kinematics, with projectile motion being the foremost example of kinematics in two-dimensions. Question 3 has 3 subparts, and covers dynamics and Newton's laws, force of gravity, force of tension, normal force, frictional forces, static and kinetic friction, and acceleration. Question 4 is the True or False section, and covers general concepts based on this unit such as physical quantities, units and dimensions, conversion factors, scalar and vector quantities, average and instantaneous velocity.

In appendix C, we see that unit test 2 also follows the same pattern in the syllabus. Question 1 has 3 subparts, and tests knowledge on circular motion, Newton's universal gravitational law, centripetal acceleration, orbital speed, orbital period, and Kepler's law. Question 2 has 4 subparts, and covers topics such as potential energy, kinetic energy, work, and conservation of energy. Question 3 has 2 subparts, and covers linear momentum, elastic, and inelastic collisions. Question 4 is the True or False section, and covers general concepts based on this unit, such as potential energy, kinetic energy, energy conservation, Kepler's law, Newton's gravitational inverse square law of force, circular and elliptical orbits, orbital period, coefficient of restitution, linear momentum, elastic and inelastic collisions.

In appendix D, we see that unit test 3 also follows the same pattern in the syllabus. Question 1 has 5 subparts, and tests knowledge on angular quantities such as period, frequency, angular velocity, moment of inertia, rotational kinetic energy, and conservation of angular momentum. Question 2 has 4 subparts, and covers topics such as static equilibrium, force, torque, stress, strain, elastic modulus, and tensile strength. Question 3 has 3 subparts, and covers Pascal's principle as applied to hydraulic systems, mechanical advantage, Archimedes principle, Bernoulli's principle, aerodynamic lift, and maximum takeoff weight. Question 4 is the True or False section, and covers general concepts based on this unit, such as torques, forces, static equilibrium, oscillations and waves, amplitude, period, frequency, wavelength, speed, energy transported by a wave, free and damped oscillations, sound intensity, sound level, electromagnetic waves, oscillation of a mass-spring system, spring stiffness, and Doppler effect.

We remark here that testing a wide variety of concepts in a single one-hour test is impossible to accomplish if the test is not designed efficiently. The design we have presented here enables flow of information within a given question. Having a single question on a central topic per page makes it easier for the students to the tackle the problem in a coherent manner. This design is beneficial over a traditional design where questions are distributed haphazardly with little or no coherence, as discussed in the next section.

### III. Advantages of the design

In this section, we discuss some of the advantages the efficient design format of a one-hour unit test has over a traditional design format:

- The design of the test clearly reflects the chapters covered in a given unit.
- A wide variety of topics covering the entire unit can be tested within an hour, so major topics are not left out.
- The design consisting of both quantitative and conceptual problems gives student the ability to enhance problem-solving and critical thinking skills.
- The exam can be completed within one hour, almost 80% of the students complete the exam within 50 minutes.
- The design is unique: same design is used for test 1, test 2, and test 3, with each test consisting of 4 questions with subparts.
- Each unit test is 4 pages long, with one question per page. This allows the student to focus on one central topic at a given time, so problems can be solved in a very coherent manner.
- Using a binary (True or False) system instead of multiple-choice system with four choices in the conceptual section reduces the amount of time spent in this section.
- The design facilitates grading**:** Grading becomes very easy and efficient. The student presents their steps, then box their final answer in the answer space provided. This makes it very convenient for the instructor. During grading, the instructor checks to make sure answer is correct. If the answer is incorrect, the instructors goes over the steps to identify areas where mistakes have been made. Partial credits are awarded accordingly based on the gravity of the mistake – is it a lack of understanding of fundamental principles or an algebraic problem. On the average, it takes about 6 minutes or less to grade one student test. This means for a typically class size of about 30 students, it only takes about 3 hours for the instructor to grade the entire test.
- Many students have made very positive comments on effectiveness of the unit test design format when they submit comments sheet as part of student evaluation or feedback about the course design.
- Figure 1 shows average one-hour unit test scores using the traditional design and the efficient design formats. We observe that on the average, test scores for the tradition design format is 76% while for the efficient design, it is 79%.

| Term | Test Type | Average Score (%) |
|---|---|---|
| Fall 2014 | Traditional | 71.5 |
| Spring 2015 | Traditional | 74.2 |
| Fall 2015 | Traditional | 73.7 |
| Spring 2016 | Traditional | 84.0 |
| Fall 2016 | Efficient Design | 78.1 |
| Spring 2017 | Efficient Design | 76.0 |
| Fall 2017 | Efficient Design | 83.4 |
| Spring 2018 | Efficient Design | 77.0 |

**Figure 1. Average test score for the traditional and efficient design test types. The average unit test score increased from 76% using traditional one-hour test exams to 79% using the efficient design format.**

This represents an increase of 3% which is quite significant. We attribute this increase to the efficient design format of the unit test, which promotes coherence in the flow of information.

### IV. Conclusion

In summary, we have presented an efficient design of a one-hour unit test for an introductory physics course. We remark here that testing a wide variety of concepts in a single one-hour test is impossible to accomplish if the test is not designed efficiently. The design we have presented here enables flow of information within a given question. Having a single question on a central topic per page makes it easier for the students to the tackle the problem in a coherent manner. This design is beneficial over a design where questions are distributed haphazardly with little or no coherence. The design makes it possible to assess both problem-solving skills via quantitative problems, and critical thinking skills via conceptual problems. The design also facilitates grading. This design should be used for any introductory level as well as upper-level physical science courses. The coherent nature in which the questions are presented could be exploited even for use in advanced graduate-level courses or for exams in which more than one-hour is allowed. We encourage the instructor to adapt the design to suit their syllabus, while maintaining the coherence, as well as other features unique to the design.

**Acknowledgement**

We acknowledge support from the department of physics, Pittsburg State University.

**Appendix A: Introductory Physics Syllabus for Spring 2018**

**Unit 1**
**Chapter 1**: Preliminaries, measurements, uncertainty, units, significant figures
**Chapter 2**: One-dimensional Kinematics
**Chapter 3**: Two-dimensional Kinematics, projectile motion
**Chapter 4**: Dynamics, Newton's Laws of Motion
**Unit 2**
**Chapter 5**: Circular Motion
**Chapter 6**: Newtonian Gravitation
**Chapter 7**: Work and Energy
**Chapter 8**: Conservation of Energy
**Chapter 9**: Linear Momentum, Momentum Conservation, Center of Mass
**Chapter 10**: Rotational Motion
**Chapter 11**: Angular Momentum, Moment of Inertia, Torque
**Unit 3**
**Chapter 12**: Statics, Elasticity
**Chapter 13**: Fluids, Pressure, Pascal's Principle, Archimedes' Principle, Bernoulli's principle
**Chapter 14**: Simple Harmonic Motion, simple pendulum
**Chapter 15**: Waves
**Chapter 16**: Sound intensity, beats, Doppler Effect
**Unit 4**
**Chapter 17**: Thermal Equilibrium, Temperature, Ideal Gas Laws
**Chapter 18**: Kinetic Theory of Gases, Molecular Speeds
**Chapter 19**: Heat and First Law, Specific heat, Latent heat, calorimetry, heat transfer
**Chapter 20\***: Second Law, Heat Engines (will be covered if time permits)

**Appendix B: Sample Unit Test 1 for Introductory Physics**

Name:_______________________________________________________________________

**PHYS 104 Test 1: Answer all questions. Show all necessary steps in your calculation. Unless otherwise stated, numerical calculations should be reported in 3 decimal places**

Useful equations:

$$v_x = v_{x0} + a_x t \qquad x = x_0 + v_{x0}t + \frac{1}{2}a_x t^2 \qquad v_x^2 = v_{x0}^2 + 2a_x(x - x_0)$$

$$v_y = v_{y0} + a_y t \qquad y = y_0 + v_{y0}t + \frac{1}{2}a_y t^2 \qquad v_y^2 = v_{y0}^2 + 2a_y(y - y_0)$$

$$\text{If } Ax^2 + Bx + C = 0 \quad \text{then} \quad x = \frac{-B \pm \sqrt{B^2 - 4AC}}{2A} \qquad \sum \vec{F} = m\vec{a} \qquad g = 9.81 \text{ m/s}^2$$

Volume of Cylinder = $\pi R^2 H$ $\qquad \pi = 3.142 \qquad F_{fr} = \mu_k F_N \qquad F_{fr} = \mu_s F_N$

1. The radius and height of a cylinder are measured as: R = 3.34 cm ± 0.05 cm and H = 7.7 cm ± 0.1 cm.

   a) **(5pts)** Calculate the % uncertainty in R.

   Answer:___________________

   b) **(5pts)** Calculate the % uncertainty in H.

   Answer:___________________

   c) **(10pts)** Estimate the % uncertainty in the volume.

   Answer:___________________

   d) **(10pts)** An object of mass 21 grams has a volume of 2.414 cm³. To the correct number of significant figures, determine the object's density in kg/m³.

   Answer:___________________

2. (30pts) A projectile is shot from the edge of a cliff 7 m above the ground level with an initial speed of 11.0 m/s at an angle of 40° with the horizontal, as shown in the figure. The target, which is a circle of diameter 3.0 m is placed so that the horizontal distance from the center of the target to the projectile is 17.0 m. Will the projectile hit the target?

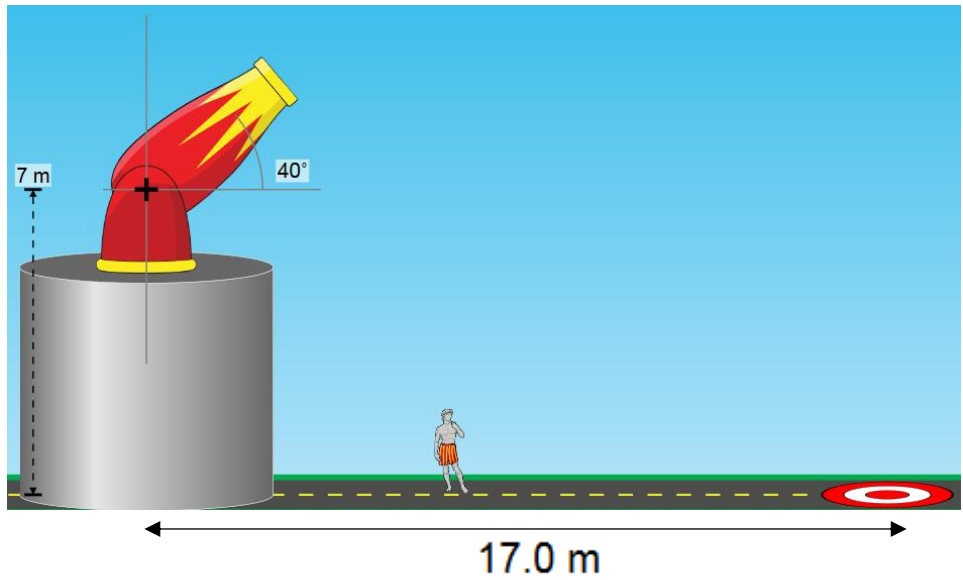

3. A **28.0 kg** block is connected to an empty **3.00 kg** bucket by a cord running over a frictionless pulley (as shown below). The *coefficient of static friction* between the table and the block is **0.50** and the *coefficient of kinetic friction* between the table and the block is **0.36**. Sand is gradually added to the bucket until the system just begins to move.

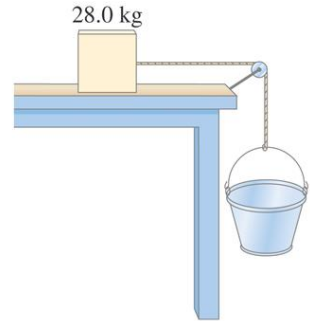

a) **(10pts)** Calculate the mass of the sand added.

**Answer:** _____________________

b) **(10 pts)** Calculate the acceleration of the system when the system starts moving.

**Answer:** _____________________

c) **(5pts)** Calculate the magnitude of the force of tension acting on the cord when the system is moving.

**Answer:** _____________________

**4. Answer True (T) or False (F): (2pts each for a total of 20 points)**

a) 36 km/h is equivalent to 10 m/s :__________________

b) Density is a fundamental physical quantity:__________________

c) The weight of an object is the force of gravity acting on the object and is measured in kilograms:__________________

d) When driving, your speedometer indicates a speed of 40 miles per hours, this is your average speed __________________

e) The standard SI unit for speed is miles per hour: __________________

f) The coefficient of static friction is always less than the coefficient of kinetic friction:________

g) Force is a vector quantity:__________________

h) The ISS refers to "International Science Station":__________________

i) You walk 3 miles to the east, and then 5 miles to the west, your net displacement is 2 miles in the east direction:____________

j) The dimension for velocity is $L/T^2$:__________________

**Appendix C: Sample Unit Test 2 for Introductory Physics**

Name:_____________________________________________________________________________

**PHYS 104 Test 2: Answer all questions. Show all necessary steps in your calculation. Unless otherwise stated, numerical calculations should be reported in 3 decimal places**

---

1. **A satellite is orbiting at a height of about 650 km above the earth's surface. The mass of the earth is $5.976 \times 10^{24}$ kg and the radius of the earth is $6.378 \times 10^6$ m.**
   a) Assuming a circular orbit, what is the speed of the satellite in its orbit? (**10pts**)

   **Orbital Speed (m/s) =** _______________________

   b) Assuming a circular orbit, what is the period of the satellite's orbit? (**10pts**)

   **Orbital Period (minutes) =** _______________________

   c) Planet A and planet B are in circular orbits around a distant star. Planet A is 49 times farther from the star than planet B. What is the ratio of their speeds $v_A/v_B$? (**10pts**)

   **Ratio of Speeds $v_A/v_B$ =** _______________________

2. **An object of mass 7.5 kg is at rest at the top of a frictionless inclined plane of length 12.0 m and angle of inclination 40.0° with the horizontal. The object is released from this position and it stops at a distance x from the bottom of the inclined plane along the horizontal surface, as shown in the figure below. The coefficient of kinetic friction for the horizontal surface is 0.250.**

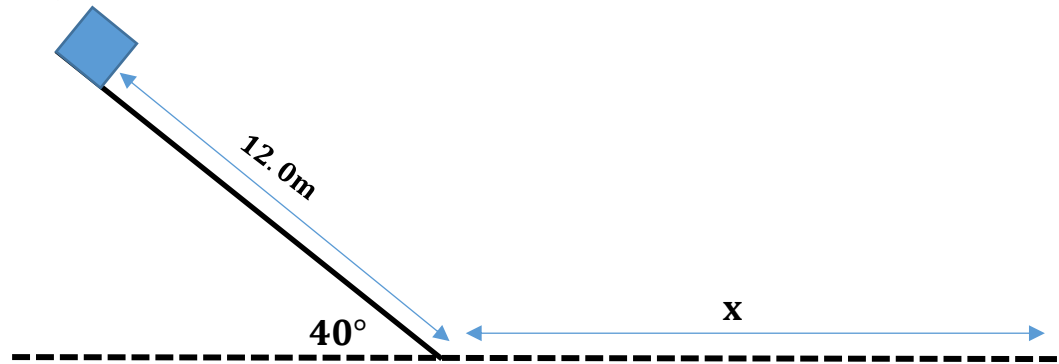

a) What is the kinetic energy of the object at the bottom of the incline plane? **(10pts)**

Kinetic Energy = ______________________

b) What is the speed of the object at the bottom of the inclined plane? **(10pts)**

Speed at bottom = ______________________

c) At what horizontal distance x from the bottom of the inclined plane will the object stop? **(10pts)**

Horizontal distance x = ______________________

3. **A 15.0 kg sphere moving in the positive x direction at 6.5 m/s collides head-on with a 10.0 kg sphere moving in the negative x direction at 5.0 m/s.**

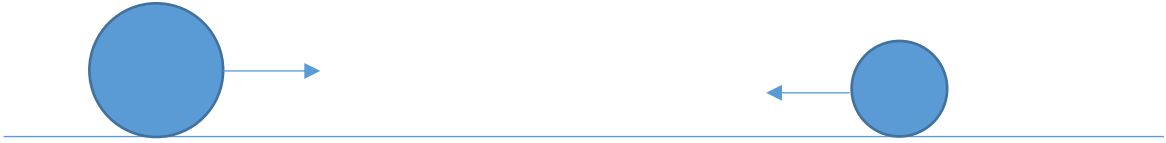

a) Calculate the final velocity of each sphere if the spheres stick together (inelastic collision). **(10pts)**

**Final velocity of 15 kg sphere** = \_\_\_\_\_\_\_\_\_\_\_; **Final velocity of 10 kg sphere** = \_\_\_\_\_\_\_\_\_\_

b) Calculate the final velocity of each spehere if the collision is elastic. **(15pts)**

**Final velocity of 15 kg sphere** = \_\_\_\_\_\_\_\_\_\_\_; **Final velocity of 10 kg sphere** = \_\_\_\_\_\_\_\_\_\_\_

4. **Answer True (T) or False (F): (2pts each for a total of 20 points)**

FIGURE 8-2

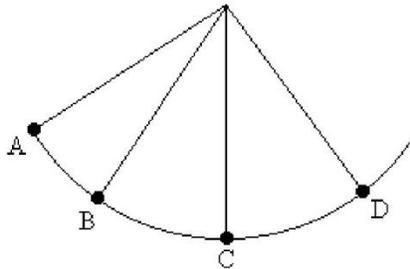

**A mass is attached to one end of a string. The other end of the string is attached to a rigid support. The mass is released at A and swings in a vertical arc to points B, C, and D.**

a) Refer to Fig. 8.2, the mass has the highest potential energy at point D:_____________________

b) Refer to Fig. 8.2, the mass has the highest kinetic energy at point C:_____________________

c) A planet has two small satellites in circular orbits around the planet. The first satellite has a period 20.0 hours and an orbital radius $2.00 \times 10^7$ m. The second planet has an orbital radius $3.00 \times 10^7$ m. The orbital period of the second satellite is equal to 10.87 hours:_____________________

d) If the distance between the center of Earth and a satellite is doubled, the magnitude of the Earth's gravitational force on the satellite decreases by a factor of 4:_____________________

e) A tennis ball of mass 65 grams is released from rest at a height of 3.0 m above the floor level. After bouncing off the floor, it rises to a final height of 1.0 m. The energy absorbed by the floor is equal to 1275.3 J:_____________

f) The orbital period of Neptune about the Sun is greater than the orbital period of Earth around the sun:_____________________

g) A tennis ball of mass 65 grams is dropped from rest at some height above the ground. It stikes the floor with a velocity of 2.0 m/s and bounces off the floor with a velocity of -2.0 m/s. We can conclude that the kinetic energy of the ball is conserved:_____________________

h) A tennis ball of mass 65 grams is dropped from rest at some height above the ground. It stikes the floor with a velocity of 2.0 m/s and bounces off the floor with a velocity of -2.0 m/s. We can conclude that the momentum of the ball is conserved:_____________________

i) Planets closer to the Sun, like Mercury, Venus, and Earth have more of a circular orbit than an elliptical orbit:_____________________

j) A force on a particle depends on position such that $F(x) = (3.00 \text{ N/m}^2)x^2 + (4.00 \text{ N/m})x$ for a particle constrained to move along the x-axis. The work done by this force on a particle that moves from x = 0.00 m to x = 3.00 m is equal to 45 Joules:_____________________

## Appendix D: Sample Unit Test 3 for Introductory Physics

Name:_______________________________________________

PHYS 104 Test 3: Answer all questions. Show all necessary steps in your calculation.

### Question One
A rotating system consisting of four blades is shown below. Each of the rotor blades can be considered a **thin rod** of length 4.5 m long and has a mass of 140 kg. Assume that it takes 0.75 seconds for the system to make one complete revolution about the axis of rotation.

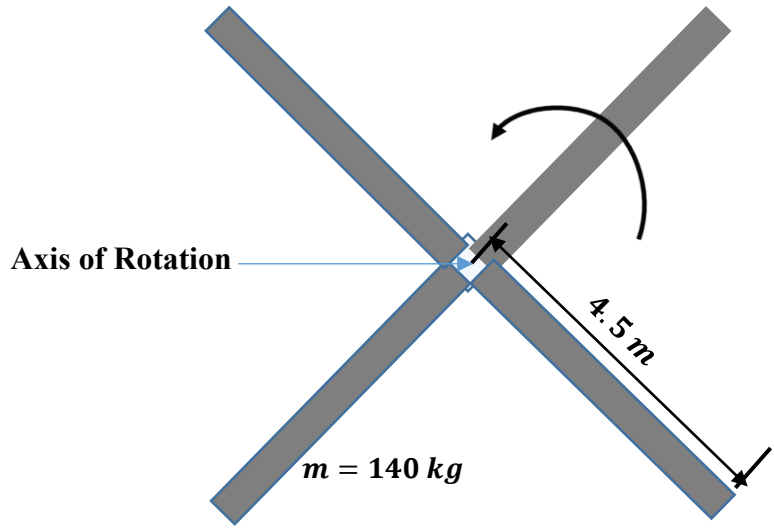

a)   What is the period (in seconds) of the system?**(5pts)**

   **Answer:** _______________________

b)   What is the angular velocity (in rad/s) of the system?**(5pts)**

   **Answer:** _______________________

c)   What is the moment of inertia (in kg m²) of the system about the axis of rotation?**(5pts)**

   **Answer:** _______________________

d)   What is the rotational kinetic energy (in Joules) of the system?**(5pts)**

   **Answer:** _______________________

A non-rotating square plate of mass 540 kg and length 4.5 m is dropped onto the freely spinning blades so that their centers are superimposed.

e)   What is the angular velocity (in rad/s) of the combined system?**(10pts)**

   **Answer:**_______________________

### Question Two
A traffic light hangs from a pole as shown in the figure below. The uniform aluminum pole AB is 8.0 m long and has a mass of 13.0 kg. The mass of the traffic light is 25.0 kg.
a) What is the tension (in Newton) in the horizontal massless cable CD?**(10pts)**

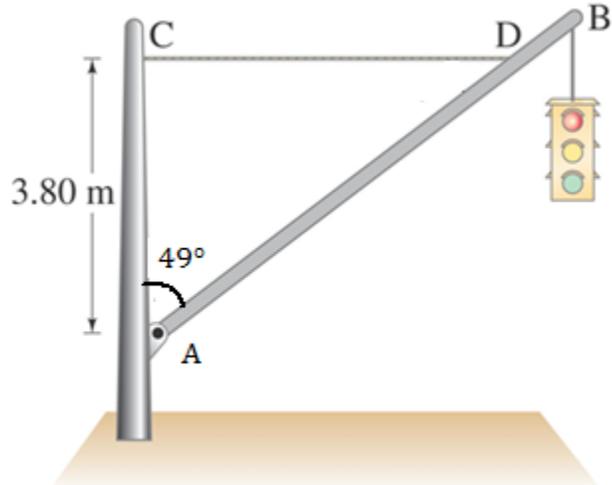

   **Answer:**___________________________

b) What is the vertical and horizontal components of the force (in Newton) exerted by the pivot A on the aluminum pole?**(10pts)**

   **Answer:**

   **Vertical Component:**_______________     **Horizontal Component:**___________________

c) The horizontal cable CD is made of steel and has a radius of 0.10 cm. What is the stress on the cable? Will it break? **Note:** Ultimate tensile stress of steel = 500 x $10^6$ N/m² **(10pts)**

   **Answer**___________________

**Question Three**

a) A hydraulic system has two pistons with diameters 6 cm and 36 cm. A force of 750 N is applied to the 6 cm diameter piston. What output force is produced at the 36 cm piston? What is the mechanical advantage of this hydraulic system? (**10pts**)

      Output Force______________________                Mechanical Advantage_______________

a) When a crown of mass 14.7 kg is submerged in water, an accurate scale reads only 13.4 kg. Is the crown made of gold? (**10pts**)

                                                                                                                                 Answer:______________________

b) A Boeing 787 Dreamliner airplane has a maximum takeoff weight (MTOW) of $2.28 \times 10^5$ kg. At takeoff, the speed of air molecules over the top and bottom surfaces of the wings are estimated to be 135 m/s and 80 m/s, respectively. Each wing of the airplane has an area of 162.5 m². Calculate the total force of lift produced by the wings. Is the force of lift sufficient? (**10pts**)

                                                                                                                                   Answer:______________________

**Question Four: Answer True (T) or False (F): (2pts each for a total of 20 points)**

**FIGURE 1**     **FIGURE 2**

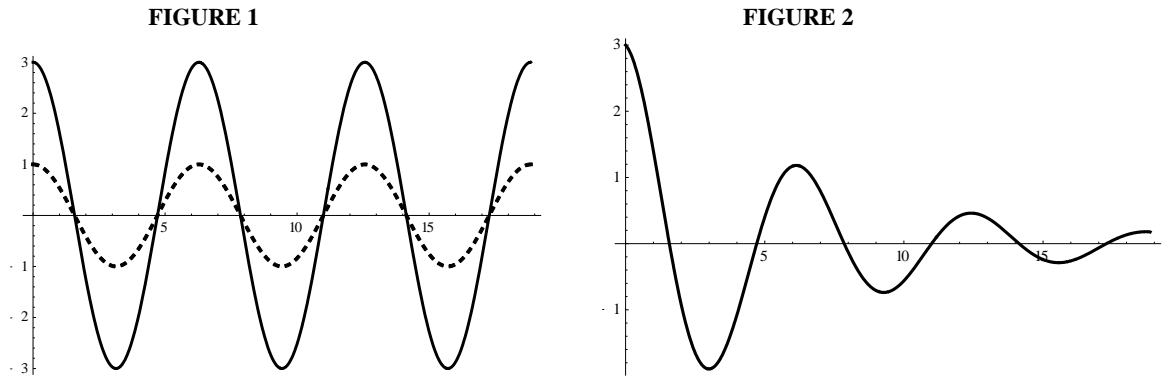

**Figures 1 and 2 show the displacement of a wave. On both figures, the y-axis represents displacement, while the x-axis represents time.**

a) Refer to Figure 1, the wave represented by the solid line curve has the same period as the wave represented by the dashed line curve:____________________

b) Refer to Figure 1, the wave represented by the solid curve has a frequency three times larger than the wave represented by the dashed line curve:____________________

c) Refer to Figure 1, the energy transported by the solid line curve wave is 9 times larger than the energy transported by the dashed line curve wave:____________________

d) Refer to Figure 1, both solid and dashed line waves describe free oscillations of the wave systems:__________

e) Refer to Figure 2, the curve represents a wave traveling in a medium in which frictional forces are present:__________________

f) The sound intensity measured 30 m away from a jet plane is 100 W/m². The corresponding sound level in decibels is 140 dB:__________________

g) In telecommunications and wireless technology, x-rays are used to transmit information because x-rays are electromagnetic waves that travel at the speed of light:______________________

h) A vertical spring of length 60 cm is hanging freely. A 100 gram mass is attached to the spring, and the final length of the spring is measured as 75 cm. The spring stiffness (k-value) is 65.4 N/m:________________

i) The light spectrum produced by a distant star was analyzed using a spectrometer over different time periods. The light spectrum reveals that the light emitted by the star is shifted more and more towards the blue end of the electromagnetic spectrum. From this information, we can deduce that the star is moving towards the earth:______________________

j) The system shown in the figure below, when released from rest will swing in the counter clockwise direction______________

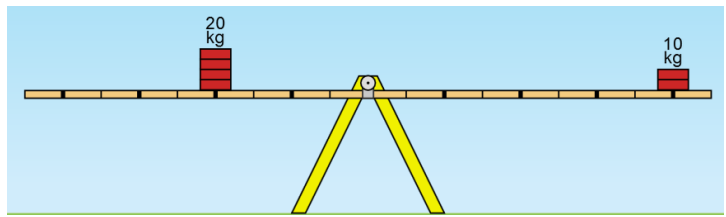